\documentclass[11pt,reqno]{amsart}

\usepackage{amsmath,amsfonts}
\usepackage{latexsym,amssymb}
\usepackage{bm}
\usepackage[dvipsnames]{xcolor}

\newcommand{\dd}{\mathrm{d}}

\begin{document}

\title{Physical meaning of electromagnetic mass and 4/3--problem}

\author{Vladimir~Onoochin}

\begin{abstract}
In this article one aspect of the so--called \lq{}4/3--problem\rq{} is analyzed, {\it namely} definitions of the electromagnetic mass of the classical electron.

It is shown that if the special relativity definition of the electromagnetic (EM) mass as the ratio of the electromagnetic field energy to the square of the speed of light is correctly treated by the scientists who considered this problem, the second definition, which originated with Thomson, {\it i.e.} a coefficient of proportionality  of the EM momentum of the particle and its velocity has another physical meaning. This meaning was explained by Frenkel in his textbook on classical electrodynamics~\cite{Fr}. According to this scientist, the second EM mass is actually a self-inductance of the classical electron or the reaction of its magnetic field to a change in the velocity of this particle.

Consequently, these two physical quantities have different meanings, and attempts to reduce the expression for one mass to an expression for the second mass have always been unsuccessful.
\end{abstract}

\maketitle

The problem of electromagnetic mass of a classical charge frequently attracts the attention of physicists (see, for example, the recent E-print by Exirifard et al.~\cite{Exir}). Such interest in this problem is caused by a certain discrepancy between two quantities associated with the so-called `electromagnetic masses' of the classical electron~\cite{Schw}. These quantities cannot be measured directly, but can only be calculated. The existing discrepancy could be ignored since it is not be verified experimentally, but physicists were concerned about the fact that this case, according to Feynman, `was a great confusion' (of the special relativity)~\cite{Fey}. 

The concept of the electromagnetic origin of the mass of the classical electron appeared at the end of XIX century when it was found that the energy of the electromagnetic field surrounding the electron arises with the arising velocity of this charge~\cite{Th}. Unfortunately, the idea to explain the whole mass of the electron by its electromagnetic origin failed. But as some experiments were made to determine the dependence of the mass (or inertia) of the accelerated electron on its velocity $v$ (the experiments of Kaufmann and Bucherer), some theoretical works appeared in the beginning of 1900$^{th}$ in attempts to describe the correct dependence $m(v)$.

In 1900, Poincar\'e derived the expression for the momentum stored in the electromagnetic fields created by a charge (in Gauss units),
\begin{equation}
{\bf P}=\frac{1}{4\pi c}\int \left[{\bf E}\times{\bf H} \right] dV\,.\label{Pem}
\end{equation}
Using this expression Abraham calculated electromagnetic mass of free electron. His result for this mass of  the electron, when its velocity tends to zero and then the 'longitudinal' and 'transverse' masses coincide, is (below Eq.~(16c) of~\cite{Abr})
\begin{equation}
m_{em}^{(1)}=\left\{
\begin{array}{ll}
\dfrac{4}{5}\dfrac{e^2}{c^2 a}\,\,\mbox{for a volume charge}\\
\\
\dfrac{2}{3}\dfrac{e^2}{c^2 a}\,\,\mbox{for a surface charge}
\end{array}\right.\,,
\end{equation}
where $a$ is the particle radius. It should be noted that the numerical estimate of this quantity was not important, since it could not be measured in experiments. Physicists were most interested in how an electron behaves as its velocity $v$ approaches the speed of light, or how and why its electromagnetic mass arises as $v\,\to\,c$. In other words, physicists were trying to understand what the physical reason for the appearance of the electromagnetic mass of charged particles is. 

The special relativity gave simple explanation of this fact (\S 10 of~\cite{AE}). According to this theory there is a connection between the mass of some physical object and its energy.

It was reasonable to suggest that such a connection exists between the energy of the EM field and its mass. So when a classical charge is at rest, this connection is of the form $\mathcal{E}=m_{em}c^2$. Since the electrostatic energy of the classical electron is known, the EM mass of this field is
\begin{equation}
m_{em}^{(2)}=\frac{1}{2c^2}\int E_0^2\dd {\bf r}=\frac{e^2}{2c^2a}\,.
\end{equation}
It was a result that puzzles the physicists since it was expected that values of two masses, $m_{em}^{(1)}$ and $m_{em}^{(2)}$ will be equal. 

This puzzle is known as the \lq{}4/3--problem\rq{}. Since the 1900$^{th}$, many attempts have been made to solve it. However, none of these attempts yielded a satisfactory solution. For example, there are statements that Poincar\'e resolved this problem in his last paper on relativity~\cite{Poinc}. However, Poincar\'e was not seeking a solution to the \lq{}4/3--problem\rq{}. His goal was to find an explanation for the stability of the electron as a charged particle by introducing a hypothetical pressure that would reign the electron. There were followers of Poincar\'e who attributed the solution of the paradox to this scientist. 

What Poincar\'e states is that this pressure is {\it negative}. However, this does not imply that the energy created by this factor is also negative. The latter is needed to reduce the value of $m_{em}^{(1)}$ to the mass determined from the electrostatic energy. If $\mathcal{E}_{non}=-\mathcal{E}_0/3$
\[
\frac{4}{3}\frac{\mathcal{E}_0}{c^2}+\frac{\mathcal{E}_{non}}{c^2}=
\frac{4}{3}\frac{\mathcal{E}_0}{c^2}-\frac{1}{3}\frac{\mathcal{E}_0}{c^2}=m_{em}^{(2)}
\]
and \lq{}the troublesome factor 4/3 has disappeared\rq{}~\cite{Jans}.

Let us repeat that the negative sign of the pressure does not mean that the internal energy of the electron is negative, and Poincar\'e did not make any statements that this energy can be negative. It would be useful to give one argument. When some capacitor is charged, the charges on its plates create negative pressure which tends to attract the plates. But the electric energy accumulated by this capacitor is strictly positive. The same should be true for Poincar\'e's pressure and its energy.

Now the dominant explanation of this paradox is that  the energy and momentum of the EM field created by the classical electron do not form a four-vector. According to~\cite{Bett},  it was Fermi who first used this to resolve the paradox. However, the Fermi paper went largely unnoticed, and the problem remained in the community. It was solved again in 1936 by W. Wilson~\cite{Wils} who added negative term to the energy of the EM field surrounding the electron. Since correctness of introducing this term is difficult to justify, the physicists returned to the idea first stated by Fermi - the  energy and momentum of the EM fields created by a classical electron are not Lorentz-invariant (comment to Eq.~(5) of~\cite{rohr}). 

Let us show that it is not so and a unsolvability of this problem is caused by other factors, {\it namely}, by incorrect treatment of the quantity associated with the electromagnetic mass which was initially introduced by Thomson. But first it is useful to demonstrate that, in full accordance to the special relativity, the energy and momentum of the electromagnetic field created by the classical electron form the four--vector. For simplicity let us consider the EM field created by this charge moving in the $x$ axis with constant velocity $v$ (K frame). Then the field energy and momentum are (in the Gauss units), 
\begin{eqnarray}
P^0=\mathcal{E}=\int \frac{E^2+H^2}{8\pi}\dd V\,; \label{en}\\
P^i={\bf P}=\int \frac{\left[ {\bf E}\times {\bf H}\right]}{4\pi c} \dd V\,,\quad i=1,2,3 \,\,.\label{mom}
\end{eqnarray}
It would be convenient to distinguish longitudinal (alone $x$ axis) component $E_{\|}$ of the electric field and its transverse component $E_{\bot}$. Then Eq.~(\ref{en}) is written as
\begin{equation}
	P^0=\mathcal{E}=\int \frac{E_{\|}^2+E_{\bot}^2+H^2}{8\pi}\dd V\,. \label{enn}
\end{equation}
Obviously, the EM momentum is formed only by the transverse field components.

Let us calculate the energy of the EM field of the particle in K$_0$ frame where the electron is at rest.
\begin{equation}
\mathcal{E}_0=\int \frac{E_0^2 }{8\pi}d V =\int \frac{E_{\|,0}^2+E_{\bot,0}^2 }{8\pi}d V \,,
\end{equation}
The electric field of the moving charge is
\begin{equation}
{\bf E}_v=\frac{q}{\sqrt{1-(v/c)^2}}	\dfrac{{\bf r}}{\left[\dfrac{(x-vt)^2}{1-(v/c)^2}+y^2+z^2\right]^{3/2}} \,,
\end{equation}	
where ${\bf r}=\{(x-vt);y;z \}$. The square of the electric field is
\begin{equation}
E_v^2=\frac{q^2}{1-(v/c)^2}	\dfrac{\tilde{x}^2+y^2+z^2}{\left[\dfrac{\tilde{x}^2}{1-(v/c)^2}+y^2+z^2\right]^{3/2}} \,.
\end{equation}	
where $\tilde{x}=x-vt$. Let us introduce spherical coordinates,
\[
R=\sqrt{\dfrac{\tilde{x}^2}{1-(v/c)^2}+y^2+z^2}\,\,;\,\, y^2+z^2=R^2\sin^2\theta\,\,;\,\, dxdydz=\sqrt{1-(v/c)^2}R^2dR \sin\theta d\theta d\phi\,.
\]
Then the square of the electric field is
\begin{equation}
E_v^2=\frac{q^2}{1-(v/c)^2}	\dfrac{\left[1-(v/c)^2\right] \cos^2\theta+\sin^2\theta}{R^4} \,.
\end{equation}
It can be rewritten as
\begin{equation}
E_v^2=\frac{1}{1-(v/c)^2}\left\{\left[1-(v/c)^2\right]E_{\|,0}^2+E_{\bot,0}^2\right\}\,,
\end{equation}
since in new spherical coordinates in frame K the expressions for the electric field coincide with the expressions for the same field written in K$_0$. 

The relation between the transverse $E_{\bot}$ field and circular $H_{\phi}$ field in K frame is simple (Eq.~(11.150) of~\cite{JDJ}),
\[
H_{\phi}=\frac{v}{c}E_{\bot}\,.
\]
Therefore, the energy density of both fields is
\begin{equation}
\varepsilon =\frac{1}{1-(v/c)^2}\frac{E_v^2+H^2}{8\pi}=\frac{1}{1-(v/c)^2}\frac{\left[ (1-v^2/c^2)
E_{\|,0}^2+(1+v^2/c^2)E_{\bot,0}^2\right] }{8\pi}\,.
\end{equation}
We need to verify the relation
\begin{equation}
\mathcal{E}_0=\frac{\mathcal{E}_v-\left({\bf v}\cdot{\bf P}\right)}{\sqrt{1-(v/c)^2}}\,, \label{rel}
\end{equation}
where 
\begin{equation}
\mathcal{E}_v=\int \varepsilon  \dd V'=\sqrt{1-(v/c)^2}\int 	 \varepsilon \dd x' \dd y\dd z \,,
\end{equation}
and 
\begin{eqnarray}
{\bf P}=\frac{\sqrt{1-(v/c)^2}}{4\pi c}\int {\bf p}\dd V =\frac{1}{4\pi c\sqrt{1-(v/c)^2}}\int [E_{\bot,0}\times H_{\phi}]  \dd x' \dd y\dd z=\nonumber\\
=\frac{1}{8\pi c^2\sqrt{1-(v/c)^2}}\int 2{\bf v}E_{\bot,0}^2 \dd x' \dd y\dd z\,,
\end{eqnarray}
where ${\bf p}$  is the density of the EM momentum

So Eq. ~\eqref{rel} takes a form
\begin{equation}
\mathcal{E}_0=\frac{1}{1-(v/c)^2}\int \frac{\left[ (1-v^2/c^2)E_{\|,0}^2+(1-v^2/c^2)E_{\bot,0}^2\right] }{8\pi} \dd x' \dd y\dd z \,,=
\int \frac{\left[ E_{\|,0}^2+E_{\bot,0}^2\right] }{8\pi} \dd x' \dd y\dd z \,.
\end{equation}
It means that the energy and momentum of the EM fields determined in two inertial frames are connected by the Lorentz transformations, or they are Lorentz-invariant. 

Similar relation is given in the textbook of Jackson, Ch.~16.5~\cite{JDJ}. 
\begin{equation}
\varepsilon-({\bf v}\cdot{\bf p})=\frac{E^2+H^2}{8\pi}-
\frac{\left({\bf v}\cdot [{\bf E}\times{\bf H}]\right)}{4\pi c}=
\frac{E^2-H^2}{8\pi}\,, \label{mass}
\end{equation}
where the author uses ${\bf B}=  [{\bf v}\times{\bf E}]/c$. Because $E^2-H^2$ is invariant, the relativistic rule
\begin{equation}
\int \frac{E_0^2}{8\pi} dV = \frac{1}{\sqrt{1-(v/c)^2}} \int \frac{E^2-H^2}{8\pi} dV' \,,\label{M}
\end{equation}
should be fulfilled. 

Since the energy $\mathcal{E}$ and momentum ${\bf P}$ are correctly described by the special relativity and at the same time the paradox is not resolved, one can state: what quantities form this paradox?

The physical meaning of $m_{em}^{(2)}$ is clear - it is mass of the electrostatic energy according to the special relativity. But $m_{em}^{(1)}$ is a mass - of what?

In \S 7 of his paper of 1903, Abraham analyzes a dynamical properties of a classical electron and among some formulas he gives explicit expressions for the electrostatic energy and the magnetic energy at the limit $v\to0$ in the volume charge (when the charge is uniformly distributed inside the particle; Eq. (15e) of~\cite{Abr}),
\begin{equation*}
W_e^{vol}=\frac{3}{5}\frac{e^2}{a}\,;\,\, W_m^{vol}=\frac{4}{5}\frac{e^2v^2}{c^2a}\,.
\end{equation*}
Since according to Abraham (the argument given below Eq.~(14c)), to recalculate the same parameters for the surface charge one should multiply $W_e^{vol}$ and $W_m^{vol}$ by the factor 5/6, which gives
\begin{equation}
W_e^{surf}=\frac{1}{2}\frac{e^2}{a}\,;\,\, W_m^{surf}=\frac{2}{3}\frac{e^2v^2}{c^2a}\,.
\end{equation}
The above expressions give the explanation of the puzzle. If $W_e^{surf}$ divided by $c^2$ represents the EM mass $m_{em}^{(2)}$ of the electrostatic energy, the ratio $W_m^{surf}/c^2$ does not give a quantity describing a mass. This parameter is calculated as
\[
m_{em}^{(1)}=\lim_{v\to0}\frac{1}{v}\frac{\partial  W_m^{surf}}{\partial v}\,.
\]
So the electromagnetic masses are obtained in quite different ways. $m_{em}^{(2)}$, calculated by relativists, is obtained from the electrostatic energy and $m_{em}^{(1)}$, calculated by Abraham and Lorentz, is a reaction of the magnetic field to a change in the velocity of the charged particle. 

Obviously $m_{em}^{(1)}$ and $m_{em}^{(2)}$ are two absolutely different quantities. This explains the failure of all attempts to transform one into the other - the electrostatic energy divided by $c^2$ cannot be transformed into the coefficient before $v^2$ in the magnetic energy by any linear transformations and even by adding some terms.  

The statement that the physical meaning of $m_{em}^{(1)}$ is not a mass, as it is accepted in classical mechanics, but \lq{}self--induction\rq{} of the classical electron is explained in the textbook of Frenkel~\cite{Fr}. Moreover, this author shows how the formula for $m_{em}^{(1)}$ follows from calculation of electromotive force of induction by which the classical electron act on itself under accelerated motion.

According to~\cite{Fr}, Ch. VII, Eqs.~(16), if $\mathcal{E}_0$ ($U$ in Frenkel's notation) is the energy of electric field of the electron at rest, the results
\begin{equation}
m_{em}^{(2)}=\frac{U}{c^2}\,;\quad m_{em}^{(1)}=\frac{4}{3}\frac{U}{c^2}\,, \label{U}
\end{equation}
do not depend on a charge distribution inside the particle; only concrete value of $U$ is determined by this distribution. Therefore addition of Poincar\'e term  cannot resolve the problem; this term should be added to both $U$ in Eq.~(\ref{U}) and the difference between two electromagnetic masses remains.

One can state a question: what mass, $m_{em}^{(1)}$  or $m_{em}^{(2)}$ , enters the expression for a dependence of the mass of the electron on its velocity (the dependence that is checked in the experiments). Frenkel shows that it is 
\[
m(v)=\frac{m_0}{\sqrt{1 - (v/c)^2}}=\frac{1}{\sqrt{1-(v/c)^2}}\frac{4}{3}\frac{U}{c^2}\,.
\]

Let us summarize what we have.\newline
$\bullet$ Two quantities, which are associated with the EM mass, are calculated from one expression: the Lagrangian of the EM field. One quantity is calculated from the electrostatic energy term of this Lagrangian (in the limit when the velocity of the charge tends to zero) and the other quantity is calculated from the magnetic energy term. It should be noted that the methods of calculation differ for both quantities. It is clear that the physical meaning of these calculated quantities must be different. It is incorrect to equate these quantities with each other.\newline
$\bullet$  In a case when one ignores the difference in physical meanings of these quantities and tries to make Lorentz transformations to convert an expression for one mass to the expression describing the other mass, these attempts are initially doomed to failure. The expression for $m_{em}^{(2)}$ is formed by three components of the electric field: two transverse components and one longitudinal component. The expression for $m_{em}^{(1)}$ is formed by two transverse components; the longitudinal component does not contribute to this "mass." It is impossible to transform a longitudinal EM field into a transverse field using relativistic transformations.

These two arguments prove beyond any doubt that there is no connection between the two quantities that are considered to be the electromagnetic masses of the classical electron. Therefore, the \lq{}4/3--problem\rq{} cannot be solved.

This problem of the coexistence of two different electromagnetic masses is not so valuable for modern physics. But one aspect of it has a certain significance for physics  - what mass is measured in the experiments to verify $m(v)$--dependence?
When some particle is being accelerated to ultra-relativistic velocity and its mass is greatly increasing, does this mean that it is not the mass itself that arises, but the surrounding magnetic field is greatly increasing in such a way that it behaves like to provide a large value of inertia, corresponding to the inertia of very heavy particles?

\end{document}